\documentclass[fleqn,usenatbib]{mnras}

\usepackage{newtxtext,newtxmath}
\usepackage[T1]{fontenc}

\DeclareRobustCommand{\VAN}[3]{#2}
\let\VANthebibliography\thebibliography
\def\thebibliography{\DeclareRobustCommand{\VAN}[3]{##3}\VANthebibliography}

\usepackage{graphicx}	% Including figure files
\usepackage{amsmath}	% Advanced maths commands
\usepackage{pdflscape}
\usepackage{threeparttable}
\newcommand\kms{$\mbox{km s}^{-1}$}

\title[Red supergiant stars in binary systems. I.]{Red supergiant stars in binary systems. I. Identification and characterisation in the Small Magellanic Cloud from the UVIT ultraviolet imaging survey}

\author[L. R. Patrick et al.]{L. R. Patrick,$^{1, 2}$\thanks{E-mail: lee@lrp-astro.org (LRP)}
D. Thilker,$^{3}$
D. J. Lennon,$^{4, 5}$
L. Bianchi,$^{3}$
A. Schootemeijer,$^{6, 7}$ \newauthor
R. Dorda,$^{3, 4}$
N. Langer$^{6, 7}$
and I. Negueruela$^{1}$
\\
$^{1}$Departamento de F\'{\i}sica Aplicada, Universidad de Alicante, E-03690 San Vicente del Raspeig, Alicante, Spain\\
$^{2}$School of Physical Sciences, The Open University, Walton Hall, Milton Keynes MK7 6AA, UK\\
$^{3}$Department of Physics and Astronomy, Johns Hopkins University, 3400 N. Charles Street, Baltimore, MD 21218, USA\\
$^{4}$Instituto de Astrof\'isica de Canarias, E-38205 La Laguna, Tenerife, Spain\\
$^{5}$Universidad de La Laguna, Dpto. Astrof\'isica, E-38206 La Laguna, Tenerife, Spain\\
$^{6}$Argelander-Institut f\"ur Astronomie, Universit\"at Bonn, Auf dem H\"ugel 71, 53121, Bonn, Germany\\
$^{7}$Max-Planck-Institut für Radioastronomie, Auf dem H\"ugel 69, 53121, Bonn, Germany
}

\date{Accepted XXX. Received YYY; in original form ZZZ}

\pubyear{2022}

\begin{document}
\label{firstpage}
\pagerange{\pageref{firstpage}--\pageref{lastpage}}
\maketitle

\begin{abstract}
We aim to identify and characterise binary systems containing red supergiant (RSG) stars in the Small Magellanic Cloud (SMC) using a newly available ultraviolet (UV) point source catalogue obtained using the Ultraviolet Imaging Telescope (UVIT) on board AstroSat.
We select a sample of 560 SMC RSGs based on photometric and spectroscopic observations at optical wavelengths and cross-match this with the far-UV point source catalogue using the UVIT F172M filter, finding 88 matches down to $m_{F172M}=20.3$\,ABmag, which we interpret as hot companions to the RSGs.
Stellar parameters (luminosities, effective temperatures and masses) for both components in all 88 binary systems are determined and we find mass distributions in the ranges $6.1 < M/M_\odot < 22.3$ for RSGs and 3.7\,$<M/M_\odot <$\,15.6 for their companions.
The most massive RSG binary system in the SMC has a combined mass of 32\,$\pm$\,4\,M$_\odot$, with a mass ratio ($q$) of 0.92.
By simulating observing biases, we find an intrinsic multipliciy fraction of $18.8\,\pm\,1.5\,\%$ for mass ratios in the range $0.3 < q < 1.0$ and orbital periods approximately in the range $3 < \log P [\rm days] < 8$.
By comparing our results with those of a similar mass on the main-sequence, we determine the fraction of single stars to be $\sim$20\,\% and argue that the orbital period distribution declines rapidly beyond $\log P \sim 3.5$.
We study the mass-ratio distribution of RSG binary systems and find that a uniform distribution best describes the data below 14\,M$_\odot$.
Above 15\,M$_\odot$, we find a lack of high mass-ratio systems.
% This is a simple template for authors to write new MNRAS papers.
% The abstract should briefly describe the aims, methods, and main results of the paper.
% It should be a single paragraph not more than 250 words (200 words for Letters).
% No references should appear in the abstract.
\end{abstract}


%%%%%%%%%%%%%%%%%%%%%%%%%%%%%%%%%%%%%%%%%%%%%%%%%%

%%%%%%%%%%%%%%%%% BODY OF PAPER %%%%%%%%%%%%%%%%%%

\section{Introduction}

It is now clear that most massive stars reside in binary or higher order multiple systems~\citep[e.g.][]{1998AJ....115..821M,2001A&A...368..122G,2014ApJS..215...15S,2017ApJS..230...15M}, with $\sim$70\,\% of close binary systems expected to
interact during their lifetimes~\citep{2012Sci...337..444S, 2014ApJS..213...34K}. 
These interactions have profound effects on the evolution of the stars in such systems~\citep{2013ApJ...764..166D} and
the nature of their subsequent supernova explosions~\citep{1992ApJ...391..246P, 2017PASA...34....1D}, as well as on the formation of stellar mass double compact object (DCO) binaries~\citep{2017A&A...604A..55M}.
The first steps are already being taken to examine how multiplicity affects the evolution of stellar populations~\citep{2008MNRAS.384.1109E, 2020ApJ...888L..12W}. Such simulations are also beginning to produce estimates of the binary properties of their evolved products that include DCOs~\citep{2020A&A...638A..39L}.

Red supergiant (RSG) stars are an important piece of this puzzle.
The vast majority of isolated massive stars (above 8M$_\odot$) experience a RSG phase either directly before core collapse as a supernova~\citep{Smartt09} or as an intermediate phase~\citep{Groh13}.
Despite their importance, and recent observational advances (see below) there remains much work to be done to understand the multiplicity properties and, in particular, the properties of the companions of RSGs.

For the closest period massive star binary systems~\citep[those within an orbital period of less than 10\,d or an orbital separation of 0.15\,au;][]{2012Sci...337..444S,2017ApJS..230...15M}, binary evolution frequently results in interactions and stellar mergers~\citep[e.g.][]{2017ApJS..230...15M,2020A&A...638A..39L,2022A&A...659A..98S}.
The products of such mergers are observed as massive analogues of blue straggler stars in young stellar clusters~\citep[e.g.][]{2014ApJ...780..117S} and, as these stars evolve to the RSG phase, they can be observed as so-called red-straggler stars~\citep{b19}.
This is supported by the recent studies of RSGs in clusters in the Magellanic Clouds and Milky Way that suggest up to 50\% of RSGs may be the result of mergers in a previous evolutionary phase~\citep{2019MNRAS.486..266B, b19,2020A&A...635A..29P}.
In addition, red stragglers can, in principle, be observed within binary systems if the system was originally a hierarchical triple system. 

Binary systems with intermediate orbital periods (10--1000\,d or $\sim$0.15--3\,au) typically interact in some form and also frequently result in stellar mergers~\citep{2020A&A...638A..39L}.
One expects Roche lobe overflow within binary systems to strip the donor's envelope of would-be RSG binary systems in favour of the production of Wolf-Rayet stars~\citep{2008MNRAS.384.1109E}.
Because of this, RSG binary systems where the RSG is the primary are expected to exist in orbital configurations where the two components are sufficiently separated that the stars evolve in effective isolation.
In systems where the RSG is the secondary\footnote{i.e. not the initially more massive component of the binary system.}, the primary will have most likely evolved to produce a compact object.
Supernova explosions within binary systems likely unbind the system~\citep{2019A&A...624A..66R} and produce massive runaway and, more commonly, walkaway stars~\citep{2019A&A...624A..66R}, which are observed at early evolutionary phases~\citep[e.g.][]{2018A&A...619A..78L} and less commonly in RSGs (e.g. $\alpha$ Orionis;~\citealt{2008AJ....135.1430H}).
The relatively few systems that remain bound can be observed as massive stars with compact object companions~\citep{2020ApJ...896...32G, 2020ApJ...904..143H,2021A&A...649A.167L}, which may produce DCO binary systems~\citep{2018MNRAS.481.1908K, 2020A&A...638A..39L}.

   \begin{figure*}
   \centering
  \includegraphics[width=\linewidth, trim=2cm 6cm 2cm 0cm, clip]{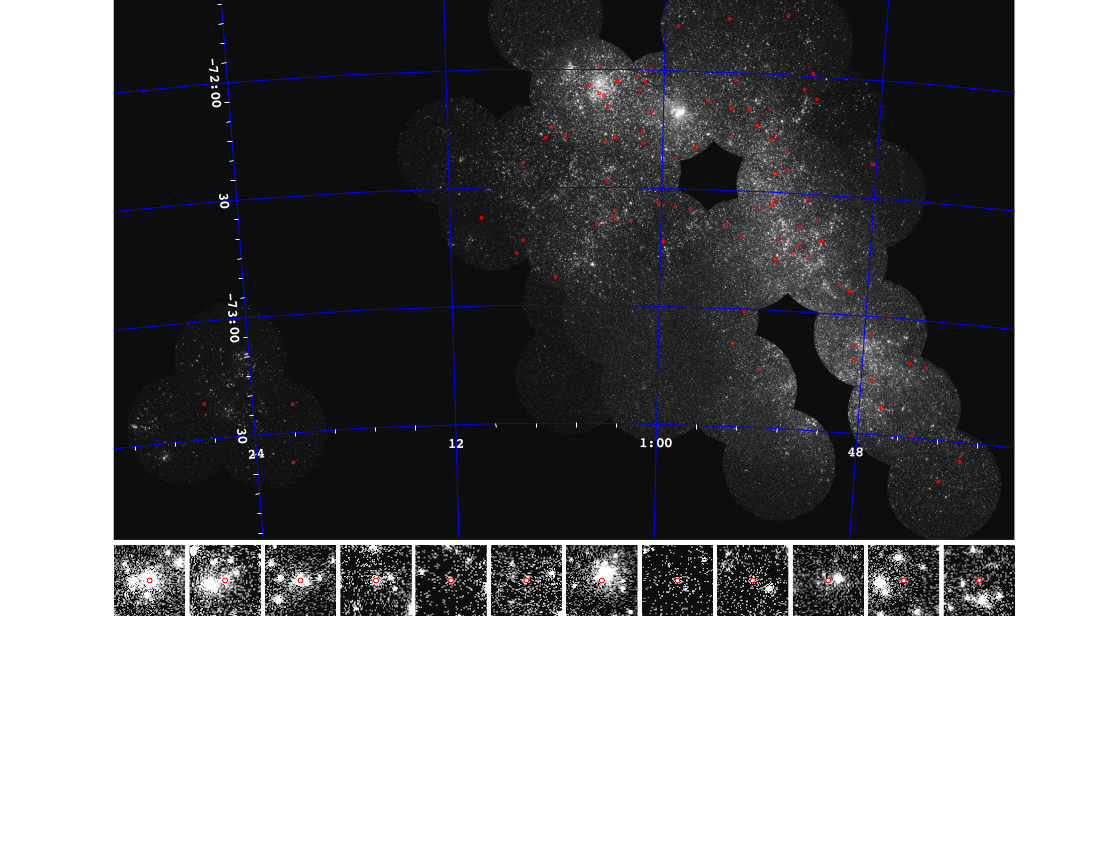}
   \caption{Mosaic of the UVIT SMC survey highlighting the locations of the matched RSG targets with red circles.
   The 12 snapshots below the main figure are examples of counterparts to RSGs in the F172M filter, which is a clear signature of a hot companion, as single RSGs will be undetected at these wavelengths.
   These snapshots are $30\times30$\arcsec\ and are ordered by the mass of the companion from the most massive on the left to the least massive on the right.
   The gaps in the main mosaic represent pending observations.}
              \label{fig:uvit-fov}
    \end{figure*}

Determination of the binary fraction of RSGs, and the nature of the companions in RSG binary systems, can therefore address a number of important issues.
The presence of a companion indicates the fraction of main-sequence progenitors that have a companion in an orbital configuration that avoids a merger event, while an independent determination of that companion's temperature and luminosity constrains the age of the system, provided the companion is a main-sequence star, resulting in an independent confirmation of whether or not the RSG is a merger product.
Characterisation of the secondary can also help uncover more exotic companions such as stripped stars, while in the case of known radial velocity variables the absence of a detectable companion can perhaps lead to the inference of black hole and neutron star companions.

Previous work in this area has focused on two methods for detecting companions: long-term radial velocity variations and detecting the presence of a hot companion in the spectrum or spectral energy distribution (SED).

Examples of the former include \citet{burki} who found a binary fraction of $\sim$35\,\% among F--M supergiants in the Milky Way, while \citet{patrick2019, 2020A&A...635A..29P} determined an upper limit of $\sim$30\,\% for RSGs in clusters in the Large and Small Magellanic Clouds (LMC and SMC respectively).
\citet[][hereafter, DP21]{2021MNRAS.502.4890D} have identified 45 Magellanic Cloud RSG binary systems using multi-epoch radial velocity information, which offers a rare opportunity to characterise systems containing a RSG that may ultimately result in DCO binary systems.

\citet{2018AJ....155..207N} developed a method to detect RSG binary systems using optical colours, which has been applied to Local Group galaxies in subsequent studies~\citep{2019ApJ...875..124N,2020ApJ...900..118N,2021ApJ...908...87N}.
These authors identified candidate RSG binary systems using photometry and follow-up spectroscopy is used to confirm the binary nature of candidates by identifying signatures of hot stars in the RSG spectra.
Using a k-nearest neighbour algorithm, and combining optical and ultraviolet (UV) photometry \citet{2020ApJ...900..118N} determined the intrinsic binary fraction of the Large Magellanic Cloud, by accounting for observational biases, to be 19.5$^{+7.6}_{-6.7}$\,\%.
In the metal rich environments of M31 and M33,~\citet{2021ApJ...908...87N} apply a similar technique to that of ~\citet{2020ApJ...900..118N} and find an intrinsic binary fraction of up to 41.2$^{+12.0}_{-7.3}$\,\% and 33.5$^{+8.6}_{-5.0}$\,\% in M33 and M31, respectively.
\citet{2021ApJ...908...87N} conclude that there exists a metallicity dependence on the RSG binary fraction. In the metal-poor environment of the SMC we can directly test this hypothesis via a comparison with similar studies in more metal rich environments.

The UV domain offers an important advantage over optical studies in that the cool supergiant, despite its larger radius, is significantly fainter than a main-sequence star in the near- and far-UV (NUV and FUV, respectively).
In this spectral region, the flux from M- and K-type supergiants is dominated by line and continuum chromospheric emission \citep{carpenter1994,carpenter2014}, that is of course not accounted for by photospheric models.
However, their brightness in the FUV is still significantly fainter than that of main-sequence B-type stars.
For example, by adopting the FUV flux for $\alpha$\,Ori from the ASTRAL $HST$ spectral library \citep{astral}, we estimate $m_{F172M}=\sim$13.9\,ABmag in the Ultraviolet Imaging Telescope (UVIT) F172M filter.
As $\alpha$\,Ori has similar extinction to our sample, but has a distance modulus of 6.1, this implies an apparent magnitude of RSGs in the SMC of around $m_{F172M}=$~26--27\,ABmag, well below our detection limit and much fainter than main-sequence B-type stars (see Section~\ref{sec:obs}).

In this paper, we take advantage of a new UV survey of the SMC using the  UVIT on board the satellite AstroSat~\citep{2006AdSpR..38.2989A,2012SPIE.8443E..1NK}.
The survey and the resultant data are presented in Section~\ref{sec:obs}, while the results and conclusions are discussed in Sections~\ref{sec:dis} and \ref{sec:conclusion}, respectively.

\section{Observations}                          \label{sec:obs}

\subsection{RSG source catalogue}

The initial source catalogue of RSGs is based on that of~\citet[][hereafter  YB20]{2020A&A...639A.116Y}.
These authors constructed a RSG catalogue in the SMC principally based on five different photometric criteria from colour-magnitude diagrams (CMDs) at different wavelengths from the SMC point source catalogue of~\citet{2019A&A...629A..91Y}.
These criteria are anchored on the appearance of the RSG population within the MESA Isochrones and Stellar Tracks~\citep[MIST; ][]{2016ApJ...823..102C,2016ApJS..222....8D}.
YB20 developed a ranking system ranging from $-$1 to 5, depending on the criteria met by each source to qualify as a RSG candidate.
Ranks 4--5 flag sources with a low probability to be a RSG.
As an initial source catalogue, we select 1233 targets from YB20 with ranks between $-$1 and 3, which correspond to either targets that have a spectroscopic classification as a RSG (rank $-$1) or with at least two independent photometric classifications (ranks 0 to 3).

With this initial source catalogue, we cross-match all targets with the Gaia~EDR3 data release~\citep{2021A&A...649A...1G} using MAST Casjobs interface.
We discard 60 sources that have a combination of 3-$\sigma$ significant parallax measurement greater than zero and renormalised unit weight error (RUWE) less than 1.5.
These sources we assume to be foreground contaminants.
In addition, we use the following criteria to exclude candidates, based on the mean proper motion and dispersion of the SMC from the most recent Gaia EDR3 results~\citep{2021A&A...649A...7G}.
We discard all sources outside a 3-$\sigma$ box from the mean proper motion values centred on $\mu_{(\alpha, \delta)} = (1.7608\,\pm\,0.4472, 0.3038\,\pm\,0.6375)$.
This results in a sample of around 1000 high-probability SMC RSG candidates.
We further restrict this sample by applying a magnitude cut by requiring M$_J \leq -6$ that is roughly equivalent to $\log L/L_\odot>3.6$ or a stellar mass of $\gtrsim7\,M_\odot$, at the distance of the SMC.
Besides focusing the sample on stars that may undergo core collapse, this refinement removes sources that might be confused with lower mass asymptotic giant branch stars (AGBs), see, for example, \citet{2015A&A...578A...3G} for spectroscopic confirmation of this contamination.
Our final sample of SMC RSG candidates consists of 862 sources and is provided in electronic form as Table~\ref{tb:params_all}.

During the latter stages of this work,~\citet{2021ApJ...922..177M} published a source catalogue of 1745 SMC RSGs that is based on the appearance of RSGs in the $J-Ks$ vs. $Ks$ CMD.
The main difference between the YB20 and~\citet{2021ApJ...922..177M} catalogues is that YB20 select sources based on mid-IR data from Spitzer Enhanced Imaging Products whereas~\citet{2021ApJ...922..177M} select sources directly from the Two Micron All Sky Survey (2MASS).
The criteria used to define the sample of~\citet{2021ApJ...922..177M} are similar to the criteria of YB20 used in the $J-Ks$ vs. $Ks$ CMD, however, the~\citet{2021ApJ...922..177M} sample go fainter in $Ks$-band magnitude and the range of $J-Ks$ colours considered is typically narrower than in YB20.
In addition, at almost the same time~\citet{2021ApJ...923..232R}, published an even-more extensive list of RSGs in the SMC, with their sample containing no fewer than 2138 RSGs.
These authors base their selection on the $J-H$ vs. $H-K$. 
We follow the recommendation of YB20 and use only RSGs that are classified as such in multiple wavelength regimes, therefore we choose to retain the sample based on the YB20 classifications.
We note that 75\,\% (651) of our sample are found within the \citet{2021ApJ...922..177M} catalogue and 78\,\% (667) are found within the~\citet{2021ApJ...923..232R} catalogue.

\subsection{Cross-matching the RSG and UVIT catalogues}        \label{sub:XM}
We matched the RSG sample with photometry from the UVIT survey of Thilker et al. (in prep.).
The SMC was surveyed with UVIT using the FUV F172M filter, which has a pivot wavelength of approximately 1707\,\AA \footnote{\url{https://uvit.iiap.res.in/Instrument/Filters}}. 
Overlapping 28\arcmin\ fields were used to observe to a 5-$\sigma$ depth of $m_{F172M}\sim$20.3\,ABmag, with point spread function full-width half-maximum of approximately 1\arcsec\ and an astrometric accuracy of approximately 0.1\arcsec. A section of the survey is illustrated in Figure\,\ref{fig:uvit-fov}.

\begin{figure}
   \centering
   \includegraphics[width=\columnwidth]{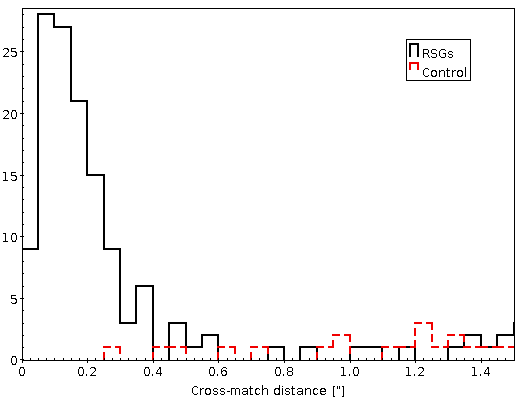}
   \includegraphics[width=\columnwidth]{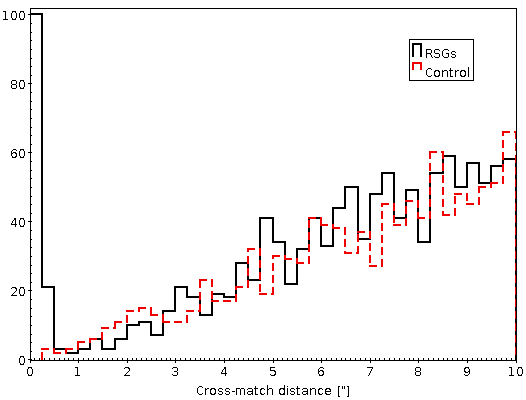}
      \caption{The separation in arcseconds between cross-matched sources from the UVIT source catalogue and the RSG source catalogue (black solid line).
      The cross-match separation of the control sample is shown with the red dashed histogram.
      See Section~\ref{sub:XM} for a detailed comparison of these two distributions.
              }
         \label{fig:XMD}
   \end{figure}

The UVIT SMC survey was completed at the $\sim$75\% level.
From the sample of 862 SMC RSG candidates, 560 lie within the footprint of the UVIT survey.
These sources are cross-matched to the UVIT source catalogue, recording all UV matches out to a cross-match distance (XMD) of 10\arcsec. 
Figure~\ref{fig:XMD} illustrates the histogram of minimum XMDs of the resulting catalogue.
From this figure, we identify a peak at a XMD consistent with zero, with a tail of this distribution extending to around $\sim$0.4\arcsec, which we interpret as genuine matches.
For XMD > 1.0\arcsec, the number of matches rises steadily and we interpret these as spurious matches. 
We find a unique match within 0.4\arcsec\ for 88 sources.
Matched UVIT sources in regions where fields overlap have more than one measurement; in these cases we choose the measurement which is the closest match to the RSG position.
The multiple measurements are consistent within their estimated uncertainty, in terms of both magnitude and position.

%
%                                                One column figure
%----------------------------------------------------------------- 

%-----------------------------------------------------------------

To quantitatively assess the impact of spurious alignments that produce false-positive results and assess the validity of our choice of maximum XMD, we repeat the cross-match process on a control catalogue.
Adapting the method of~\citet{2020ApJS..250...36B}, the control sample consists of the full RSG source catalogue offset by 20\arcsec\ in declination.
An offset of 20\arcsec\ is chosen to mimic, as closely as possible, the density and distribution of the underlying RSG source catalogue.
The results are illustrated in Figure~\ref{fig:XMD} as the red dashed histograms.
These results clearly demonstrate that false positives through chance alignments have a negligible contribution to the distribution of genuine matches (i.e. the black solid histogram at 0.1\arcsec) and that chance alignments contribute close to 100\% of the matches outside 0.4\arcsec.
The bottom panel of Figure~\ref{fig:XMD} demonstrates that, at large XMD, the two distributions are effectively identical.
This strengthens the assumption that for a XMD larger than 0.5\arcsec\ the UVIT matches to the input catalogue can be assumed to be positional coincidences, as their distribution matches that of the underlying population and their number increases geometrically with the area as the XMD increases.
The catalogue of genuine matches is provided in electronic form as Table~\ref{tb:params_bs}, together with their derived stellar parameters (see Section~\ref{sub:parameters}).

%----------------------------------------------------------------
%
%                                                One column figure
%----------------------------------------------------------------- 
   \begin{figure}
   \centering
   \includegraphics[width=\columnwidth]{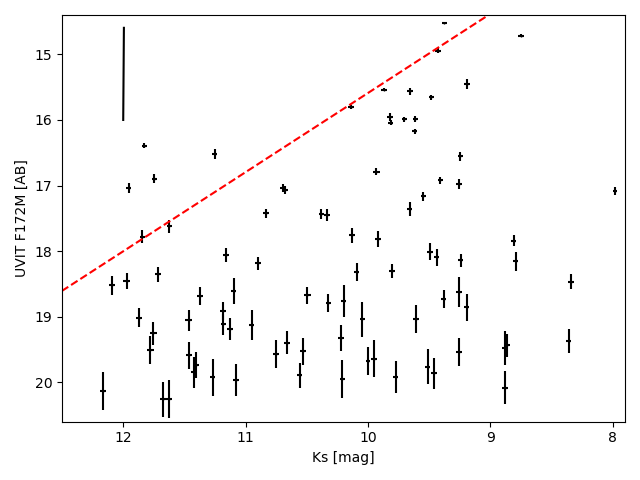}
      \caption{UVIT F172M-band magnitudes shown against $Ks$-band magnitudes for the candidate RSG binary systems.
      The red dashed line shows the UVIT magnitude at 10k\,K within stellar evolutionary models (see Section~\ref{sub:parameters} and Figure~\ref{fig:HRD-binaries}), assuming an age appropriate of a RSG with the corresponding J-band magnitude.
      Companions with UVIT magnitudes above the red dashed line are inconsistent with being located on the main sequence given the age of the RSG.
      The red dashed line is calculated using a comparison with stellar models, with the bolometric corrections from the MIST stellar tracks~\citep{2016ApJ...823..102C,2016ApJS..222....8D}.
      The solid black, almost vertical, line highlights the reddening vector assuming $A_{\rm V}=0.35$\,mag and the reddening law of~\citet{2003ApJ...594..279G}.
              }
         \label{fig:fuvvsj}
   \end{figure}
%-----------------------------------------------------------------

\section{Results and discussion}                                          \label{sec:dis}

Based on the arguments presented in the previous section, we assume the detected UVIT counterparts represent the hot companions of RSG binary systems.
Therefore, it is straightforward to determine that 15.7\,$\pm$\,1.5\,\% of our sample of RSGs in the SMC have a hot companion with a mass greater than $\sim$3.5\,$M_\odot$, assuming a UVIT F172M completeness limit of $m_{F172M}=20.3$\,ABmag and interstellar extinction of $E(B-V) = 0.13$ (see Section~\ref{sub:ext}).
Figure~\ref{fig:fuvvsj} compares the F172M-band and $Ks$-band magnitudes of the binary candidates.
We interpret this figure as a comparison of companion mass to RSG mass, as the F172M-band and $Ks$-band magnitudes can be used as proxies for the companion and RSG masses, respectively, as we show in the subsequent subsections.

\begin{figure}
   \centering
   \includegraphics[width=\hsize]{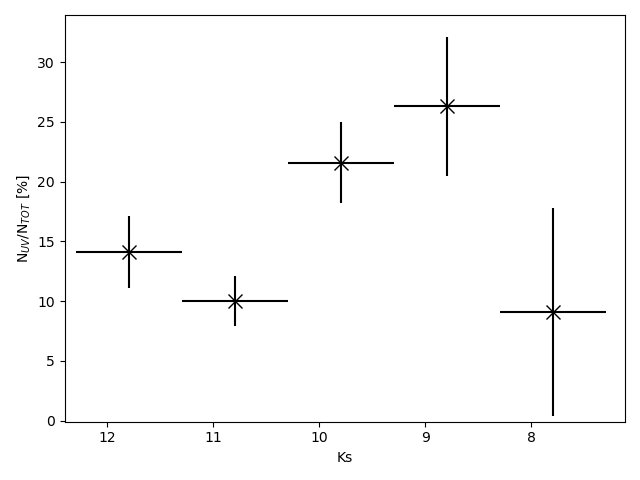}
    \caption{The observed percentage of the number of UV detections as a function of RSG magnitude. We interpret this as the binary fraction of RSGs.
    This figure illustrates the decreasing trend for fainter, and hence less massive, RSGs.
    The \textquoteleft uncertainties\textquoteright~in $Ks$-band magnitude represent the bin width.}
         \label{fig:bfvsJ}
   \end{figure}

\begin{figure*}
  \centering
  \includegraphics[width=0.8\linewidth]{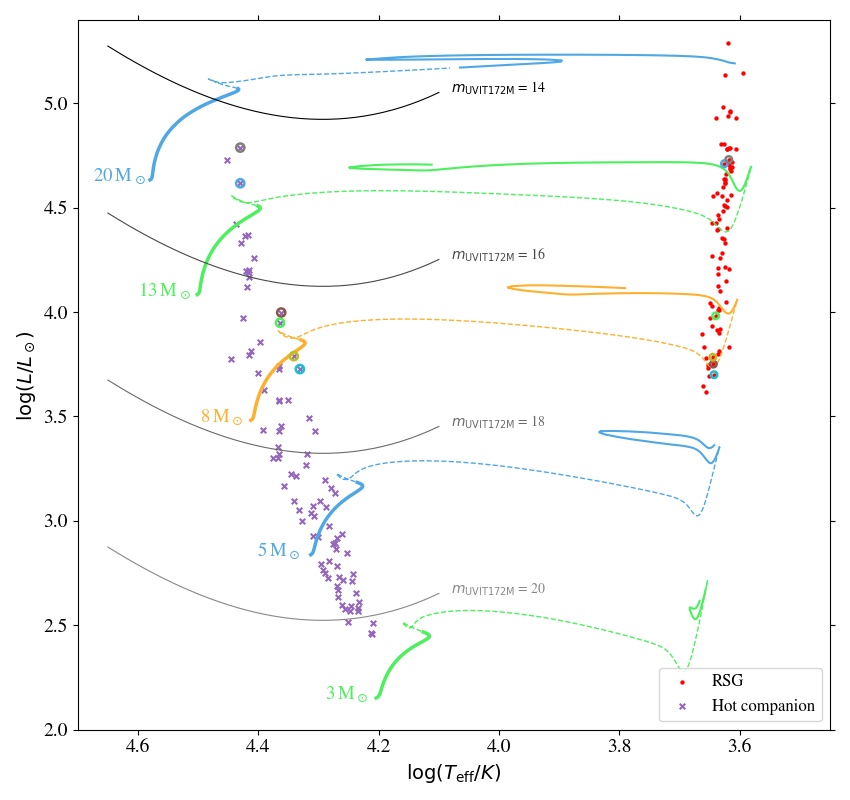}
     \caption{HRD showing both components of the RSG binary systems detected with UVIT photometry.
     Stellar evolutionary tracks are based on~\citet{2019A&A...625A.132S}.
     Solid red circles show the RSG component of the system and violet crosses show the hot companions.
     Solid grey lines show lines of constant UVIT magnitude in the stellar models.
     Rings of different colours highlight the six systems with a mass ratio greater than 1.0.
     Thick colored lines indicate core H-burning, thinner solid lines indicate core He-burning and dashed lines indicate the short phase in between.
     }
        \label{fig:HRD-binaries}
  \end{figure*}

Figure~\ref{fig:bfvsJ} demonstrates that the observed binary fraction depends on the $Ks$-band magnitude of the RSG, ranging from $\sim$0.2 for the brighter (higher mass) stars to $\sim$0.1 for the fainter (lower mass) stars.
The detection limit imposed by the UVIT photometric completeness limit ($m_{F172M}=20.3$\,ABmag or $\sim$3.5\,M$_\odot$ for a zero-age main-sequence star) results in a mass-dependent mass ratio ($q$) observing bias, such that, for the faintest RSGs in the sample, mass ratios can be detected in the range $q > 0.6$ and in the range $q > 0.3$ for the brightest RSGs in the sample. 
To quantify this, we determine stellar parameters (i.e. effective temperatures, luminosities and masses) for both components in Section~\ref{sub:parameters}, study the mass-ratio distribution in Section~\ref{sub:qs} and simulate the observing biases to determine the intrinsic multiplicity fraction in Section~\ref{sub:bf}.
Figure~\ref{fig:HRD-binaries} displays the RSG binary systems on the Hertzsprung--Russell Diagram (HRD), which allows a better visualisation of the key results of this study.

\subsection{Extinction and reddening}               \label{sub:ext}
To accurately determine stellar parameters of both components within the binary systems we must provide a consistent treatment of extinction and reddening values. 
To do this we follow~\citet{2021A&A...646A.106S} and assume a constant $A_{\rm V}=0.35$\,mag, where we use the SMC bar reddening law of~\citep{2003ApJ...594..279G} to determine extinction parameters in the FUV, for the hot companions, and near-IR, for the RSGs.

The scale of the intrinsic variations of extinction values within the SMC is typically small~\citep{1995ApJ...438..188M,1997A&A...317..871L,2002AJ....123..855Z}, but with a potentially important tail to higher extinction values.
The origin of such a tail is additional extinction in specific regions, which are attributed to specific regions, rich in hot, young, stars~\citep{2002AJ....123..855Z}.
To determine the intrinsic spread of extinction values at the location of each of our targets we use the recently published extinction maps of~\citet{2021ApJS..252...23S}.
\citet{2021ApJS..252...23S} determined $E(V-I)$ values using red clump stars from the Optical Gravitational Lensing Experiment (OGLE-IV).
For our targets, this results in an average extinction value of
$A_{\rm V}=0.17$\,mag\footnote{$A_{\rm V}$ is calculated assuming $A_I=1.5 \times E(V-I)$, as listed on the webpage associated with \citet{2021ApJS..252...23S}, and assuming the SMC bar reddening law of \citet{2003ApJ...594..279G}.} with a typical dispersion of 0.15.
We chose to retain the extinction value of $A_{\rm V}=0.35$\,mag~\citep{2021A&A...646A.106S}, rather than use the extinction values determined by \citet{2021ApJS..252...23S}, as the extinction values determined from red clump stars are more appropriate for lower mass stars.
However, we argue that the dispersion determined from these extinction maps is an accurate indicator of the local spread of extinction values for each target, given that the SMC contains little intrinsic dispersion.
Therefore, each target is assigned an uncertainty on $A_{\rm V}$ defined by the local dispersion from \citet{2021ApJS..252...23S}.
The scatter on the $A_{\rm V}$ values of our targets is correlated with larger $A_{\rm V}$ values from the ~\citet{2021ApJS..252...23S} maps, so in this sense, a small tail of higher extinction values is identified, which extends up to $A_{\rm V}=0.70$\,mag.

In previous studies of SMC RSGs a range of $A_{\rm V}$ values are assumed.
\citet{Levesque06} determined $A_{\rm V}$, effective temperatures and surface gravity values of 37 SMC RSGs by fitting spectrophotometric observations and found an average $A_{\rm V}=0.53$\,mag, with a dispersion of 0.35.
Similarly, \citet{2013ApJ...767....3D} determined $A_{\rm V}$ by fitting spectroscopic observations for 10 SMC RSGs that resulted in an average value of 0.5, with a dispersion of 0.2.
Recently,~\citet{2021MNRAS.505.4422G} updated the spectral fitting using the observations of \citet{2013ApJ...767....3D} and redetermined $A_{\rm V}$ values for the 10 SMC targets.
These authors found an average $A_{\rm V}=0.67$\,mag with a dispersion of 0.26.
In their recent study of SMC RSGs, to determine stellar parameters \citep{2021ApJ...922..177M} assumed $A_{\rm V}=0.75$\,mag to determine stellar parameters of their sample of SMC RSGs based on the results of \citet{Levesque06}.
~\citet{2018MNRAS.478.3138D} study 245 of the brightest RSGs in the SMC and determine their extinction value using the maps from the hot star sample of~\citet{2002AJ....123..855Z} obtain an average $A_{\rm V}=0.46$\,mag and a dispersion of 0.15.

Spectroscopic $A_{\rm V}$ determination is likely the most robust method to determine the extinction values of RSGs, as such estimates take into account circumstellar extinction around RSGs.
That being said, spectroscopically determined $A_{\rm V}$ values have large uncertainties and the agreement between the $A_{\rm V}$ measurements between the stars in common between the spectroscopic studies of~\citet{Levesque06} and~\citet[][]{2013ApJ...767....3D}, is poor, and has not been resolved with the updated calculations of~\citet{2021MNRAS.505.4422G}.
In particular, the stars with $A_{\rm V}>0.5$\,mag in~\citet{Levesque06}, all have $A_{\rm V}$ lower values in~\citet{2021MNRAS.505.4422G} and those with $A_{\rm V}<0.5$\,mag all have larger values in~\citet{2021MNRAS.505.4422G}.
The studies of \citet{Levesque06} and \citet{2021MNRAS.505.4422G} targeted more luminous RSGs than are present in the UVIT cross-matched sample, which are known to have additional circumstellar material and larger $A_{\rm V}$ values~\citep{2020ApJ...900..118N}.
Indeed, by measuring the infrared excess of RSGs,~\citet{2010AJ....140..416B} demonstrate that, in general, the population of SMC RSGs is relatively dust-free, which suggests that the impact of circumstellar material is low for the majority of SMC RSGs.
In this respect, we argue that $A_{\rm V}=0.35$\,mag remains a robust average value for the stars in our sample.
We comment on the impact of our choice of extinction in the following sections.

\subsection{Stellar parameters}                     \label{sub:parameters}

\subsubsection{RSG masses}

While RSG masses are controversial 
\citep{davies2020,2020MNRAS.494L..53F,2021A&ARv..29....4S}, much of this discussion centres on the masses of RSGs just prior to core collapse.
Our interest, however, is on the evolutionary phase that is typical of our RSG population and hence we use the mid-point of the RSG core He-burning phase as representative of typical properties (age and mass) of the RSG as a function of luminosity.
For this purpose, we use the MESA models~\citep{2011ApJS..192....3P,2013ApJS..208....4P,2015ApJS..220...15P,2018ApJS..234...34P} computed for the SMC based on an extension of the model grids published by~\citet{2019A&A...625A.132S}, which use a mass-dependent convective overshooting parameter ($\alpha_{\rm ov}$) and a semi-convection parameter ($\alpha_{\rm sc}$) of 10, described in~\citet{2021A&A...653A.144H}.

To determine RSG masses, we first determine their luminosities using the calibration RSG luminosity of ~\citet{2013ApJ...767....3D} with de-reddened $Ks$-band photometry.
$Ks$-band photometry has the advantage of minimising the effects of interstellar and circumstellar extinction over other near-IR photometry.
For example, we find that luminosities determined using the $Ks$-band are slightly systematically larger than those determined using the $J$-band for the brightest stars in our sample.
This trend towards larger luminosities in the $Ks$-band increases as a function of luminosity and reaches 0.10 dex at a $\log L/L_\odot = 5.1$, which is potentially the result of an increase in circumstellar extinction around higher-luminosity RSGs.
For these calculations we assume a distance modulus to the SMC of 18.95~\citep{2014ApJ...780...59G} and the $Ks$-band magnitudes are de-reddened assuming $A_{\rm V}=0.35$\,mag and the SMC bar reddening law from~\citet{2003ApJ...594..279G}.
The choice of $A_{\rm V}$ has a small impact on the RSG luminosities as a result of the shape of the adopted reddening law.
The uncertainties on the measured luminosities are typically $\pm$0.14\,dex.
Using a comparison to the models we determine a relationship between the luminosity and evolutionary mass of SMC RSGs.

For the RSG sources with UVIT counterparts, we determine masses in the range $6.2 < M/M_\odot < 20.3$, as shown in Figure~\ref{fig:HRD-binaries}. 
The most massive RSG with a UVIT counterpart is LHA\,115-S~30 with a mass of 20.3\,M$_\odot$ and a companion mass of 9.1\,M$_\odot$.
For comparison, the most massive RSG in the initial sample has a mass of 25.2\,M$_\odot$.

In addition, we determine the effective temperature for the targets. 
The RSG effective temperature scale, particularly at low-metallicity, remains uncertain.
There exists several photometric and spectroscopic techniques to determine RSG effective temperatures.
To determine the effective temperatures for the RSGs we use a calibration of near-infrared (near-IR) photometry for SMC RSGs published in DP21.
This calibration is based on the RSG effective temperature measurements of \citet[][henceforth TDN18]{2018MNRAS.476.3106T}.
TDN18 determined stellar parameters, including effective temperatures, for over 150 SMC RSGs by fitting a selection of well-separated atomic features from medium-resolution spectra in the Calcium triplet (CaT) region ($\sim$8500\,\AA) with grids of MARCS~\citep{2008A&A...486..951G} and KURUCZ~\citep{2012AJ....144..120M} one-dimensional atmospheric models, under the assumption of Local Thermodynamic Equilibrium.
We further discuss photometric and spectroscopic effective temperature determinations for SMC RSGs in Appendix~\ref{ap:Teff}.

\subsubsection{Companion masses}

To determine the masses of the companions, we assume that each companion is a single, main-sequence star that is coeval with the RSG, thus occupying the same isochrone.
For each companion, we determine the relationship in the stellar models between luminosity and effective temperature given the UVIT magnitude.
The luminosity of the companion is determined using the equation:

\begin{equation}
    \log L = (M_{bol,\odot} - m_{F172M} - BC(T_{\rm eff}) + A_{F172M} + \mu)/2.5
\end{equation}

\noindent where $\mu$ is the SMC distance modulus, namely $\mu=18.95$~\citep[][as for the RSG companions]{2014ApJ...780...59G}, and $A_{F172M}$ is the extinction in the UVIT~F172M filter, which is defined as $A_{F172M} =A_{\rm V}\times 4.013$ for the SMC bar~\citep{2003ApJ...594..279G}, where $A_{\rm V}=0.35$\,mag~\citep{2021A&A...646A.106S}.
While a tailored approach for each star may lead to more precise estimates of the ambient extinction, for example, taking account of detailed extinction maps or estimates of SMC and MW contributions to the total extinction, we estimate that the uncertainties introduced by neglecting such effects are well within our adopted uncertainties of $\pm0.14$ dex in $\log L/L_\odot$.
The bolometric correction ($BC$) is taken from the MIST models~\citep{2016ApJ...823..102C,2016ApJS..222....8D}.
In the MIST models, the bolometric correction of the F172M filter is only very weakly dependent on surface gravity and, as such, we use $\log g = 3.0$ and we assume it is solely a function of effective temperature in the temperature range studied.
$m_{F172M}$ is the apparent magnitude in the UVIT~F172M filter and we use $M_{bol,\odot}=4.74$.
The solid grey lines in Figure~\ref{fig:HRD-binaries} illustrate how the stellar parameters vary as a function of UVIT magnitude; note that these lines are almost perpendicular to the main sequence, which allows for a precise determination of stellar parameters at a given age, despite the observational limitation of only a single photometric filter.
The intersection between the isochrone, defined by the age of the RSG, and the line of constant UVIT magnitude are used to determine the mass and effective temperature of the companion.

Given the shape of the constant UVIT magnitude curves in the MIST models, for the low-mass companions the determined mass does not strongly depend on the age assumption.
With increasing companion mass, the age assumption becomes more important and, as a result, the most massive companions in the sample are those that are potentially most affected by uncertainties in the RSG age.
The choice of the extinction for the SMC is guided by the results of~\citet{2021A&A...646A.106S}, see Section~\ref{sub:ext}. 
A fixed extinction value for all of our sources neglects the effects of variable and circum-binary extinction.
Additional extinction would act to increase the companion masses determined, whereas the RSG mass estimate is not strongly dependent on the adopted extinction value.
As with the age assumption, the highest mass companions are those that are most affected by the choice of the extinction value. 
To improve the determination of the stellar parameters of the companion star requires additional UV photometric or spectroscopic observations.

No coeval solutions on the main sequence were found for six systems; in these cases the derived age is the age of a terminal-age main-sequence (TAMS) model that has the observed UVIT magnitude.
These six systems are identified in Figure~\ref{fig:fuvvsj} as systems that lie above and to the left of the red dashed line and also in Figure~\ref{fig:HRD-binaries} as coloured rings.
Because the age of the system is only approximated by the mid-point of the helium burning phase, the ages of more evolved RSGs in binary systems will tend to be underestimated, which could allow for solutions on the main sequence.
While this approach is clearly approximate, for low-mass secondaries it is reasonably accurate, since evolution on the main sequence has little impact on the F172M-band magnitude in these stars.
Further, the approach does not take into account scenarios in which the secondary is not a single, coeval main-sequence star, as will be discussed in Section~\ref{sub:qs}.
The uncertainties on the masses are dominated by the spread in the extinction law (propagated from the luminosities).

We find companion masses in the range 3.7\,$<M/M_\odot <$\,15.6, and these stars are shown in the HRD in Figure~\ref{fig:HRD-binaries}.
The most massive companion is Dachs SMC 1-13 with a mass of 15.6\,$\pm$\,0.1\,M$_\odot$.

   \begin{figure}
   \centering
   \includegraphics[width=\hsize]{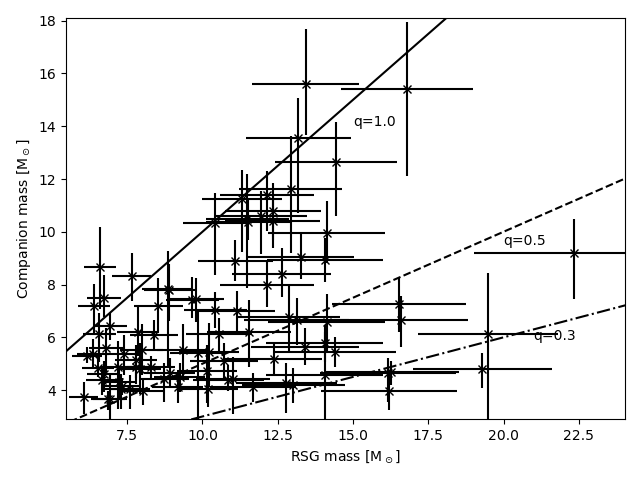}
      \caption{Derived masses for both components of the systems with UVIT counterparts.
      Straight lines highlight lines of constant mass ratio ($q$).
              }
         \label{fig:masses}
   \end{figure}

\subsection{Mass ratios of the binaries}                     \label{sub:qs}

\begin{figure}
  \centering
  \includegraphics[width=\hsize]{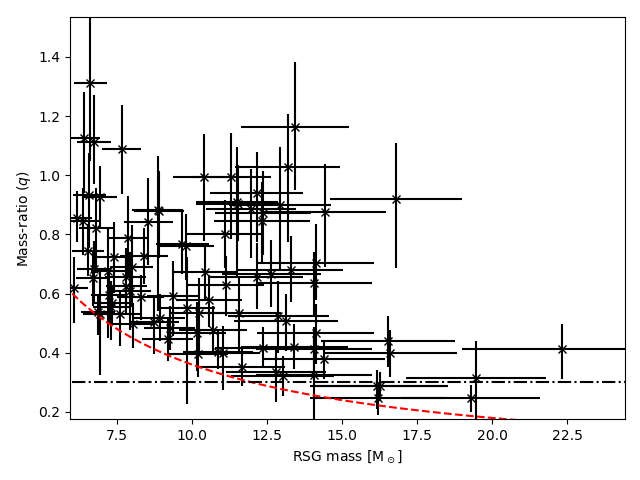}
    \caption{Mass-ratio distribution as a function of RSG mass. The dashed line indicates the detection threshold, showing how the low-$q$ systems are missing for the lower mass RSGs. The dash-dot line highlights $q=0.3$, which is the limit of the observing bias correction in Section~\ref{sub:bf}.}
         \label{fig:qvsMrsg}
  \end{figure}

The comparison of RSG and companion masses is shown in Figure~\ref{fig:masses}, while the distribution of mass ratios as a function of RSG mass is shown in Figure~\ref{fig:qvsMrsg}. 
The latter illustrates that the detection limit for the secondaries ($\sim$3.5\,$M_\odot$) leads to a progressive loss of low-mass-ratio systems with decreasing RSG mass.
As discussed at the beginning of this section, this affects the observed binary fractions, which range from $\sim$20--25\% for the higher masses to $\sim$10--15\% for the lowest mass bins (see Figure~\ref{fig:bfvsJ}).
When analysing the mass ratios distribution to minimise the effect of variable observing biases, we consider the 32 RSGs in the 10 to 14\,$M_\odot$ mass range and find that, to a good approximation, the distribution of mass ratios is flat in the range $0.3 < q < 1.0$.

These figures also show a dearth of $q>0.5$ systems for RSG masses of greater than 15\,M$_\odot$:
eight out of nine RSGs with masses above 15\,M$_\odot$ have mass ratios less than 0.5. 
This is striking, as the high-mass-ratio systems should, in principle, be the most easily detected systems in our survey. 
Indeed, by simulating the observed population of RSG binaries, we find that in 99.9\% of 10\,000 simulations more than two systems are detected in this parameter space, assuming a flat mass-ratio distribution.
Therefore, we conclude that this dearth of high-mass-ratio systems above 15\,M$_\odot$ is not the result of small number statistics at high masses, but the result of a rapid change in the observed mass-ratio distribution function.
To describe the distribution of observations above 15\,M$_\odot$ requires a power-law mass-ratio distribution function: $p_q \propto q^\kappa$, where $\kappa \sim -2$.
A tendency towards low-mass companions is expected for long-period systems~\citep{2017ApJS..230...15M}; however, such a pronounced change at 15\,M$_\odot$ is not expected. 
This is perhaps a hint that the most massive RSGs are mainly the result of stellar mergers, and the extant binaries that we observe are therefore triple systems with a low mass tertiary, while the inner binary has merged.
Of course, such a system would appear younger that its actual age \citep{b19} and also result in the overestimation of secondary masses, which further exacerbates the lack of high $q$ systems.
An alternative explanation for such a deficit is an underestimate of the extinction values.
Many authors have demonstrated the effects of increased circumstellar extinction around higher-luminosity RSGs~\citep[e.g.][for SMC RSGs]{2010AJ....140..416B}.
If we assume  hat such stars have twice the nominal extinction value (i.e. $A_{\rm V}=0.7$\,mag), we find around 50\% of the binaries have $q > 0.5$, which is potential indicator that larger extinction values may be more appropriate for this mass range.

We identify six systems with $q > 1$.
In addition to the evolved binary scenario considered in the previous section, systems with a more complex evolutionary history may be the explanation of the observed systems at $q > 1$.
The mass ratios presented in this section have been determined with the assumption that the secondary is a single, coeval, main-sequence star.
Other possibilities exist, of course, and while we cannot distinguish between these cases with our limited photometric data, we consider the impact of some of these alternative scenarios.
If the secondary is the product of a previous binary interaction, the above approach would be inappropriate.
For example, stripped stars \citep{2018A&A...615A..78G} with initial masses above about 3.5\,$M_\odot$ would fall within our magnitude range.
Additional UV data would be required to characterise the FUV sources more accurately, by either extending the UV spectral energy distribution or acquisition of UV spectra. For the latter, an exploratory snapshot programme is underway with the Hubble Space Telescope (GO\,16776, PI: Patrick).
If the companion star is the result of a previous binary merger, the above approach would be equally inappropriate.
If the hot companion is an unresolved binary, the effect would also be to overestimate the mass of an individual component.
Decreasing the $A_{\rm V}$ value for these targets would improve the situation. We note that most of the RSGs with $q > 1$ have low RSG masses.
If we assume the $A_{\rm V}$ value of~\citet{2021ApJS..252...23S} for these objects, we find only one RSG binary system with $q > 1$, namely SSTISAGEMA J005145.35-723114.8.

\subsection{Intrinsic multiplicity fraction and comparison with previous studies}          \label{sub:bf}

\begin{figure}
  \centering
  \includegraphics[width=\hsize]{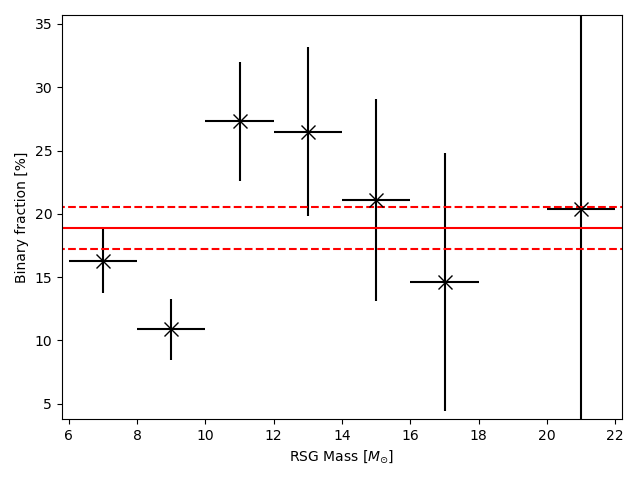}
    \caption{The intrinsic RSG binary fraction of the SMC a function of RSG mass. The red solid line shows the observing bias-corrected binary fraction for the entire RSG sample for $q > 0.3$.}
         \label{fig:ibfvsM1}
  \end{figure}

To accurately compare with previous studies, we must clearly define the parameter space over which these observations are sensitive and take into account observational biases where possible.
The principal observing bias in this study stems from the UVIT photometric completeness limit, which results in a mass ratio bias that is a function of primary mass (as illustrated in Figure~\ref{fig:qvsMrsg}).
We account for this bias by simulating the observations assuming a flat mass-ratio distribution in the range $0.3 < q < 1.0$ and a total sample size of between 90 and 125 binary systems.
These samples are defined by drawing binary systems from the initial RSG source catalogue at random and assigning mass-ratios assuming a flat mass-ratio distribution.
For each drawn sample, we determine the number of systems above and below the observing detection limit and compare this with the observed sample.
From 100\,000 simulations with randomly drawn sample sizes between 90 and 125, we find $\sim$3500 with 88 systems detected above the observing limit.
From the distribution of these samples we determine that, on average, 17.7\,$\pm$\,4.6 systems lie below our detection limit.
Taking this into account results in an intrinsic multiplicity fraction of SMC RSGs of $18.8\,\pm\,1.5\,\%$, over a range of $0.3 < q < 1.0$.
The intrinsic multiplicity fraction as a function of RSG mass is shown in Figure~\ref{fig:ibfvsM1}.
We observe the same trend in the bias-corrected multiplicity fraction as is observed in the observed multiplicity fraction such that RSGs below $\sim$10\,M$_\odot$ have a multiplicity fraction of $\sim$12.5\% and RSGs above $\sim$10\,M$_\odot$ have a binary fraction closer to $\sim$25\%.
One expects the multiplicity fraction to increase as a function of primary mass~\citep{2017ApJS..230...15M}; however, such a step is unexpected and must be studied further in other environments to confirm or refute this observation.

As previously noted, the UVIT observations are effectively insensitive to orbital periods of binary systems; however, we can place limits on the orbital periods by considering the limitations of our observations.
By using a XMD of 0.4\arcsec\ between the RSG and UVIT catalogues implies an upper limit of $\log P [\rm days]\sim 8$. 
The physical size of RSGs places an approximate lower limit on the binary systems that we can detect at $\log P [\rm days]\sim 3$.
Therefore, the intrinsic multiplicity fraction calculated can be considered to be drawn from orbital periods within the range $3 < \log P [\rm days] < 8$.

DP21 used radial velocity measurements to determine a lower limit on the RSG binary fraction of the SMC to be 15\,$\pm$\,4\,\% and, while this is consistent with our results, a direct comparison is difficult as radial velocity methods are biased against long period systems and these authors were unable to account for their observational biases.
To more accurately compare these results, we determine that 75\% (226) of the 303 SMC targets in DP21 are within our initial source catalogue, of which 41 systems have a UVIT detection.
This results in a binary fraction of 18.1\,$\pm$\,2.6\% and, by accounting for the UVIT observing bias using the same method described above, an intrinsic binary fraction of 19.0\,$\pm$\,2.6\%.
This result is in good agreement with the results determined using the full UVIT sample and is reassuringly larger than the lower limit imposed by DP21.
A direct quantitative comparison is complicated, as DP21 identified 21 `reliable' binaries within their SMC sample (6.7\% of their sample), 16 of which are in our survey area, but only 10 of these are detected as FUV sources.
The remaining five systems lie outside our survey area.
The six RSG binaries with no UVIT counterparts are perhaps a hint that the number of false positives in DP21 is significant, reflecting the diverse nature of the radial velocity sources.
We discuss these further in Section~\ref{sub:BHs?}.
The 10 UVIT detected systems are highlighted in Table~\ref{tb:params_bs} and represent the most well-characterised SMC RSG binary systems.

In other radial velocity studies, \citet{2020A&A...635A..29P} used high-precision HARPS measurements of nine RSGs in the SMC cluster NGC\,330 to derive a bias-corrected binary fraction of 30\,$\pm$\,10\,\% for systems with $2.3<\log P [\rm days]<4.3$ and $q>0.1$.
Similar work by \citet{patrick2019} on 17 cluster and field RSGs in the 30 Doradus region of the LMC yielded an upper limit of 30\% on the binary fraction, with the parameter range $q>0.3$ and $3.3<\log P [\rm days]<3.5$.
Within the Milky Way, \citet{burki} studied radial velocities over a 5-year baseline using the CORAVEL survey of 181 supergiants with spectral types F--M.
These authors found an overall binary rate of $\sim$35\,\% including spectroscopic binaries (21\,\%), suspected spectroscopic binaries (4--7\,\%) and \textquoteleft very separated binaries\textquoteright~(6--10\,\%).
For the purposes of this comparison, we exclude the \textquoteleft very separated binaries\textquoteright, as their orbital configurations are likely to be quite different from the spectroscopic binaries and their numbers and bias correction are quite uncertain.
The K- and M-type supergiant samples of~\citet{burki} are perhaps closest in evolutionary phase to the current dataset and among their 84 K- and M-type supergiants they find 19 binaries, plus 7 suspected binaries, resulting in a binary rate of $\sim$30\,\%, with an estimated uncertainty of $\sim$10\,\%.

\citet[][]{2018AJ....155..207N} proposed a similar approach to that adopted here, although they searched for companions in the $U$- and $B$-bands, as opposed to the FUV.
Essentially, this method hinges on detecting a possible B-type companion in the blue part of the visible spectrum.
This method was applied to the LMC \citep{2020ApJ...900..118N} and in a refinement of the technique these authors included NUV photometry for 75\,\% of their sample from GALEX and used a spectroscopically confirmed sample of RSG+B binaries to train a k-nearest neighbour algorithm to find candidate binaries using a $UBV$ catalogue and the GALEX UV photometric catalogue.
Using this method \citet{2020ApJ...900..118N} assign a percentage likelihood of binarity to each star in the sample and determined a bias-corrected binary fraction of 19.5$^{+7.6}_{-6.7}$\,\% for the LMC, with which we find excellent agreement in the SMC.

\begin{figure}
  \centering
  \includegraphics[width=\columnwidth]{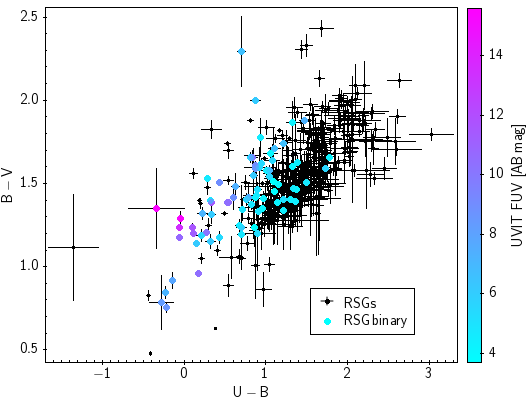}
      \caption{$U-B$, $B-V$ colour-colour diagram with photometry from~\citet{2002AJ....123..855Z}.
      Black points represent the initial target selection and coloured diamonds represent RSG binary systems with UVIT counterparts where the colour is determined by the UVIT F172M-band magnitude of the companion.
              }
         \label{fig:ubbv}
  \end{figure}

To assess potential differences between the selection criteria we cross-match our RSG sample with the SMC $UBV$ catalogue of \citet{2002AJ....123..855Z}.
We find 504 reliable matches, 75 of which are binary systems detected with UVIT photometry.
The RSG binaries have a range of $-0.2 < U-B < 1.7$, with 51 systems (68\%) having $U-B < 1.0$ (see Figure~\ref{fig:ubbv}), which is broader than the distribution of $U-B$ colours than the spectroscopically confirmed RSG binary sample in the LMC of~\citet{2020ApJ...900..118N} and in M31 and M33 of~\citet{2021ApJ...908...87N}.
Around 1/3 of RSG binaries in our sample have $U-B > 1.0$, which is significantly larger than the 8\% of stars found with $U-B > 1.0$ in the LMC sample of spectroscopically confirmed RSG binaries~\citet{2020ApJ...900..118N} and the 0\% in the samples of M31 and M33~\citep{2021ApJ...908...87N}.\footnote{The significance of these discrepancies are calculated by determining the uncertainties of the percentages assuming $\sigma = \sqrt{p(1-p)/n}$, where $p$ is the result and $n$ is the population size.}
It appears unlikely that metallicity variations between the samples can fully account for their differences, which raises the possibility that~\citet{2020ApJ...900..118N} and~\citet{2021ApJ...908...87N} might have missed RSGs binaries with fainter companions.

\citet{2020ApJ...900..118N} and \citet{2021ApJ...908...87N} determined the likelihood of chance alignments masquerading as RSG binary systems to be less than the 2\,\% level.
To test this, we experiment using a range of XMDs to determine the percentage of false positives for a given XMD, assuming that the 88 detections with 0.4\,\arcsec to be the \textquoteleft true\textquoteright~number of RSG binaries.
We find that at a XMD of 1.0" a contribution of 6\% from false positive detections, rising to 35\% at 3.0".
Figure~\ref{fig:XMD} also implies that the false positive issue for blue sources becomes a much more serious problem for more distant galaxies such as M31 or M33, however, high resolution photometry improves this situation such as the HST photometry of~\citet{2021ApJS..253...53W}, which covers around 42\,\% of the M33 sample of~\citet{2021ApJ...908...87N} and 21\,\% of their M31 sample. 

In addition, \citet{2020ApJ...900..118N} provide a preview of their upcoming SMC study where 22 SMC RSGs were observed in the same observing run as their LMC results.
Of these 22 RSGs,~\citet{2020ApJ...900..118N} classify eight as RSG binary systems based on their spectroscopic observations.
We find that 18 confirmed RSG binaries from~\citet{2020ApJ...900..118N} are within the UVIT survey footprint. Of these 18 we find 11 RSGs with UVIT counterparts, confirming six of the eight RSG binaries identified~\citet[][]{2020ApJ...900..118N}.
The remaining two systems were outside of the UVIT survey footprint, which results in an accuracy of 100\%.
The range of $U-B$ magnitudes of the 22 RSGs in \citet{2020ApJ...900..118N} extend up to $U-B = 1.0$, therefore we cannot assess the accuracy of the spectroscopic classification for $U-B > 1.0$, which is potentially where this method breaks down given the diminishing contribution from the companion.
With these observations we independently confirm the spectroscopic classification of \citet{2020ApJ...900..118N} is an effective method to identify RSG binary systems for $U-B < 1.0$.
However, almost 50\% of the stars classified as RSG binaries via UV photometry that are in the sample of \citet{2020ApJ...900..118N} were not classified as RSG binaries based on a spectroscopic identification.
These systems have a range of UVIT magnitudes of 18.8 to 16.8, corresponding to masses from 5.1 to 8.9\,$M_\odot$.
Based on these results, we conclude that the greater flux contrast enabled by FUV imaging leads to a significantly improved detection efficiency for companion masses below around $9\,M_\odot$.

Turing our attention on different evolutionary phases, a comparison with Cepheids is useful given the similarities in mass range and evolutionary state of the primary stars.
To our knowledge, no systematic study of the binary properties of SMC Cepheids exist.
\citet{2015AJ....150...13E} studied a Galactic sample of Cepheids using the radial velocity survey CORAVEL.
These authors found an observed binary fraction of 29\,$\pm$\,8\,\%, which is free of any significant observational bias, within the orbital period range $2.5 < \log P[\rm days] < 4$.
In addition, these authors found a flat mass-ratio distribution for their studied period range.
This quantitative agreement between the Galactic Cepheid and the SMC RSG binary fractions and orbital configurations suggests that metallicity does not play a primary role in the evolution of wide binary systems, supporting the conclusions of~\citet{2013ApJ...778...95M} and~\citet{2019ApJ...875...61M}.

As part of the VLT FLAMES Tarantula Survey~\citep{2011A&A...527A..50E}, \citet{2015A&A...580A..93D} determined the intrinsic binary fraction for B-type stars in the LMC.
These authors considered orbital periods in the range $0.15 < \log P [\rm days] < 3.5$ and accounted for observational biases to determine that around 60\% of B-type stars have a companion over the range $0.1 < q < 1.0$.
Taking into account the range in orbital periods studied by these authors and assuming a flat $\log P [\rm days]$ distribution over this range results in a multiplicity fraction of 20\% per decade in $\log P$ up to $\log P [\rm days] \sim 3$.
Combining this with the results presented here, which cover orbital periods outside the range considered by~\citet{2015A&A...580A..93D}, we find an intrinsic single star fraction of massive stars of $\sim20\%$, which is in good agreement with previous estimates of this statistic~\citep{2017ApJS..230...15M}.
This simple analysis also yields the important conclusion that the $\log P [\rm days]$ distribution must decline rapidly outside $\log P [\rm days] \sim 4$, which is supported by the observation of very wide OB-type binary systems~\citep{2019MNRAS.486.4098I} and may have consequences for binary formation.

In general there is reasonable consensus that the binary fraction of RSGs is approximately 20\%, with the present work finding a flat mass-ratio distribution, for $q>0.3$ and $M_{\rm RSG} <14\,M_\odot$.
As has already been noted in the literature, this is substantially less than the binary fraction of their progenitors, the OB-type stars \citep[as derived in][]{2012Sci...337..444S,2013A&A...550A.107S,2015A&A...580A..93D,2017ApJS..230...15M}, as a result of a combination of evolution and binary interaction.
The periods of RSG binaries are very long, with minimum periods of at least $10^2$--$10^3$\,d as a result of the radius of the RSG \citep{patrick2019}.
It is therefore tempting to compare our mass-ratio distribution with that for long period OB stars. 
Indeed the literature survey  of \citet{2017ApJS..230...15M} gives a mean value of the power-law exponent of the mass-ratio distribution  $\kappa = -1.7$ to $-2.0$ for long orbital period systems in the mass range covered in the present study, in clear tension with the flat distribution ($\kappa = 0.0$) found in this study below 14\,M$_\odot$.
However, there are a number of potential problems with simple comparisons of binary fractions and $q$ distributions.
For example, many OB binaries are triple, or higher order, systems whose periods may evolve, possibly driving the inner binary to merge~\citep{2021arXiv210804272T}.
On the other hand, some single RSGs may have previously been in a binary system that has since dissociated, perhaps as a result of the primary exploding as a supernova.
Therefore, this complex problem of quantitatively relating the binary properties of RSGs to their OB star ancestors will be addressed in a future binary population synthesis approach currently in preparation.

\subsection{Interesting non-detections}          \label{sub:BHs?}

One of the explanations for the known binary systems from DP21 that are not detected at UV wavelengths are compact companions.
Such systems are expected to be rare, but these systems can potentially be identified through a lack of a UV counterpart, assuming no interaction between the RSG wind and the compact companion.
Six systems have significant radial velocity variations from DP21, with no detectable hot companion more massive than 3\,M$_\odot$. 
These positive detections in DP21 are unlikely to have low-mass-ratio configurations given the scale of the detected radial velocity variability.

As stated in Section~\ref{sec:dis}, it is probable that these non-detections are the result of false positive measurements in DP21; however, because of the evolutionary significance of the detection of RSG + compact companion systems, these systems are worth noting and investigating further.
The six candidate RSG + compact companion systems are
SV*~HV2232,
Cl*~NGC~371~LE~39,
BBB~SMC~306,
PMMR~9,
Dachs~SMC~1-4 and
SV*~HV~832.
From DP21, these six systems all have radial velocity variations greater than 11.4\,\kms and between three and seven spectroscopic epochs.
SV*~HV2232 appears in the XSHOOTER Spectral library~\citep[XLS;][]{2019A&A...627A.138A} and on inspection of its spectral appearances displays no spectroscopic signatures of binarity.
To place limits on the orbital configurations that can reproduce the observations requires detailed simulations and an exhaustive search of available observations, therefore, for these reasons we reserve placing limits on potential orbital configurations for a future publication.

\section{Summary and Conclusions}                                                   \label{sec:conclusion}

In the context of the importance of multiplicity within the evolution of massive stars, the observational properties of RSGs in binary and multiple systems relatively remain poorly understood, despite recent observational progress.
In this article, we have aimed to better understand the multiplicity properties of RSGs in the SMC by identifying RSG binary systems and characterising the systems using newly available UVIT imaging.
Detecting RSG binary systems using UV photometry has a distinct advantage over other methods of detection, as it allows a direct characterisation of the companion star in a wavelength range where the RSG provides no contribution.

From a total of 560 RSGs within the UVIT survey area, 88 RSG have UVIT counterparts brighter than the limiting magnitude of approximately $m_{F172M}~=~$20.3\,ABmag. 
Based on these results, we determined the observed RSG main-sequence multiplicity fraction of the SMC to be 15.7\,$\pm$\,1.5\,\%, which can be thought of as a lower limit on the intrinsic RSG binary fraction.

Near-IR photometry was used to determine stellar parameters of the RSGs and the UVIT photometry was used to determine the stellar parameters of the companions, which assumed the same age for both components within the systems.
We used MESA models adapted from~\citet{2019A&A...625A.132S} including a mass-dependent convective overshooting parameter to compare with observations using bolometric corrections from MIST tracks~\citep{2016ApJ...823..102C,2016ApJS..222....8D}.
Figure~\ref{fig:HRD-binaries} displayed our results for both components in a HRD for the 88 RSG binary systems.

For the binary systems we found RSG masses in the range $6.2 < M_{RSG}/M_\odot < 20.3$ and companion masses in the range $3.6 < M_2/M_\odot < 15.4$ and used these results to determine the mass-ratio distribution of long-period massive binary systems, which is best described by a uniform distribution in the range $0.3 < q < 1.0$.
We found six systems that have mass ratios greater than 1.0, which are either genuine inverted mass systems or systems in which the companion represents an unresolved binary system. 
In addition, we found six candidate RSG + compact companion systems that require future study.

We simulated observational biases to determine the intrinsic multiplicity fraction of SMC RSGs and found it to be $18.8\,\pm\,1.5\,\%$, over a range of $0.3 < q < 1.0$ and $3< \log P [\rm days] < 8$.
This result is in good agreement with a lower limit set using spectroscopic observations at longer wavelengths (DP21) and those from photometric studies in other galaxies~\citep{2020ApJ...900..118N}.
Interestingly, we found a potential transition in the multiplicity fraction of RSGs at $\sim$10\,M$_\odot$, where the multiplicity fraction was lower below this value and higher above this value.
In addition, we combined our results with those at earlier evolutionary phases to estimate the single star fraction of massive stars to be $\sim20\%$.

This photometric identification of the companions of RSGs in binary systems represents the first time that the companions of a large sample of RSGs have been directly studied.
The combination of the UV and the high spatial accuracy afforded by the UVIT observations result in an accurate and precise determination of the RSG multiplicity fraction in a way that has not previously been possible given observational limitations in the Magellanic Clouds.
Follow-up Hubble Space Telescope UV spectroscopy of a sub-sample is in progress to better understand the nature of the companions and refine the stellar parameters.

\section*{Acknowledgements}
The authors would like to thank the anonymous referee for a providing useful comments that improved the quality of the article.
LRP acknowledges the support of the Generalitat Valenciana through the grant APOSTD/2020/247.
This research is partially supported by the Spanish Government under grant PGC2018-093741-B-C21 (MICIU/AEI/FEDER, UE).
DJL acknowledges support from the Spanish Government Ministerio de Ciencia, Innovaci\'on y Universidades through grants PGC-2018-091 3741-B-C22 and from the Canarian Agency for Research, Innovation and Information Society (ACIISI), of the Canary Islands Government, and the European Regional Development Fund (ERDF), under grant with reference ProID2017010115.
This work has made use of data from the European Space Agency (ESA) mission
{\it Gaia} (\url{https://www.cosmos.esa.int/gaia}), processed by the {\it Gaia}
Data Processing and Analysis Consortium (DPAC,
\url{https://www.cosmos.esa.int/web/gaia/dpac/consortium}). Funding for the DPAC
has been provided by national institutions, in particular the institutions
participating in the {\it Gaia} Multilateral Agreement.
The authors acknowledge the support of the from the Generalitat Valenciana through the grant PROMETEO/2019/041.
This is a pre-copyedited, author-produced PDF of an article accepted for publication in MNRAS following peer review. The version of record [DOI: stac1139] is available online at: xxxxxxx .

%%%%%%%%%%%%%%%%%%%%%%%%%%%%%%%%%%%%%%%%%%%%%%%%%%
\section*{Data Availability}

It is the authors' intention to make the data that have been used to determine the results and publish this article as freely and easily available as possible to permit readers to reproduce these results.
To this end, the RSG source catalogue and their derived parameters are published fully in Table~\ref{tb:params_all}.
Table~\ref{tb:params_bs} provides a list of RSG binary systems along with their derived parameters in full. 

% The stellar models that have been used to produce Figure~\ref{fig:HRD-binaries} are made publicly available in~\citet{2019A&A...625A.132S}. 

% The inclusion of a Data Availability Statement is a requirement for articles published in MNRAS. Data Availability Statements provide a standardised format for readers to understand the availability of data underlying the research results described in the article. The statement may refer to original data generated in the course of the study or to third-party data analysed in the article. The statement should describe and provide means of access, where possible, by linking to the data or providing the required accession numbers for the relevant databases or DOIs.



\bibliographystyle{mnras}
\bibliography{journals} % if your bibtex file is called example.bib

%\begin{thebibliography}{99}
%\end{thebibliography}

%%%%%%%%%%%%%%%%%%%%%%%%%%%%%%%%%%%%%%%%%%%%%%%%%%

%%%%%%%%%%%%%%%%% APPENDICES %%%%%%%%%%%%%%%%%%%%%

\appendix

\section{Tables of stellar parameters}

The following tables detail the stellar parameters for the entire RSG sample (\ref{tb:params_all}) and the RSG binary systems with a UV counterpart (\ref{tb:params_bs}).
The versions displayed in this section are printable versions of the full tables, which are made available as machine-readable versions through the VizieR database of astronomical catalogues at the Centre de Données astronomiques de Strasbourg (CDS) website and include uncertainties on all parameters for each target and positional information to aid future identifications.

% \begin{landscape}
\begin{table*}
% \centering
% \begin{threeparttable}
\caption{A sample of the RSG source catalogue with their derived parameters. A full version of this table is available as supplementary material (online).}
\label{tb:params_all}
\begin{tabular}{lcccccccc}
\hline
ID & $\alpha$ & $\delta$ & J$^a$ & Ks$^b$ & M$_1$ & err & $T_{\rm eff}^c$ & $\log L_1/L_\odot^d$\\
& & & mag & \multicolumn{2}{c}{M$_\odot$} & K \\
\hline
SSTISAGEMA J002619.90-724740.7 & 00 26 19.87419 & -72 47 40.8370 & 12.57 & 11.68 & 6.8 & 0.6 & 4330 & 3.81 \\
2MASS J00264567-7355235 & 00 26 45.67524 & -73 55 23.6052 & 11.73 & 10.67 & 8.9 & 0.8 & 4050 & 4.22 \\
SSTISAGEMA J002957.92-732228.4 & 00 29 57.91954 & -73 22 28.5285 & 11.8 & 10.89 & 8.4 & 0.8 & 4280 & 4.13 \\
SSTISAGEMA J003149.89-734351.0 & 00 31 49.89149 & -73 43 51.1320 & 12.52 & 11.67 & 6.9 & 0.6 & 4380 & 3.82 \\
SSTISAGEMA J003155.20-733442.7 & 00 31 55.20747 & -73 34 42.7821 & 12.12 & 11.26 & 7.6 & 0.7 & 4370 & 3.98 \\
2MASS J00315561-7335080 & 00 31 55.61855 & -73 35 08.0956 & 11.43 & 10.6 & 9.1 & 0.8 & 4420 & 4.24 \\
SV* HV 11223 & 00 32 01.59928 & -73 22 34.8042 & 11.18 & 9.99 & 10.9 & 1.2 & 3850 & 4.49 \\
SSTISAGEMA J003333.90-735641.7 & 00 33 33.90077 & -73 56 41.8020 & 11.49 & 10.59 & 9.1 & 0.8 & 4310 & 4.25 \\
2MASS J00335492-7319547 & 00 33 54.94717 & -73 19 54.7808 & 12.42 & 11.43 & 7.3 & 0.6 & 4160 & 3.91 \\
SSTISAGEMA J003427.70-730115.8 & 00 34 27.71514 & -73 01 15.8598 & 12.63 & 11.82 & 6.6 & 0.5 & 4460 & 3.75 \\
SSTISAGEMA J003457.93-734610.1 & 00 34 57.91440 & -73 46 10.2261 & 12.57 & 11.71 & 6.8 & 0.6 & 4370 & 3.8 \\
SkKM   2 & 00 35 03.25275 & -73 32 58.8521 & 11.59 & 10.79 & 8.6 & 0.8 & 4470 & 4.17 \\
SSTISAGEMA J003503.35-741212.1 & 00 35 03.33468 & -74 12 12.2594 & 10.95 & 10.11 & 10.5 & 1.1 & 4400 & 4.44 \\
SSTISAGEMA J003550.22-734916.5 & 00 35 50.20066 & -73 49 16.5841 & 12.26 & 11.4 & 7.3 & 0.6 & 4370 & 3.92 \\
UCAC3 33-1843 & 00 36 09.20818 & -73 43 14.9453 & 11.34 & 10.28 & 10.0 & 0.9 & 4060 & 4.37 \\

% \ldots & \ldots & \ldots & \ldots & \ldots & \ldots\\
\hline
\multicolumn{8}{l}{\footnotesize{$^a$ Photometry from The Two Micron All Sky Survey~\citep{2006AJ....131.1163S}. Typical uncertainties in range 0.020 to 0.030.}}\\
\multicolumn{8}{l}{\footnotesize{$^b$ Photometry from The Two Micron All Sky Survey~\citep{2006AJ....131.1163S}. Typical uncertainties in range 0.020 to 0.030.}}\\
\multicolumn{8}{l}{\footnotesize{$^c$ Uncertainties of $\pm 150$\,K.}}\\
\multicolumn{8}{l}{\footnotesize{$^d$ Typical uncertainties of $\pm 0.12$.}}\\
\end{tabular}
% \begin{tablenotes}
% \item[1] spacecraft will hit earth on way back since $r_p<r_{earth}$
% \end{tablenotes}
% \end{threeparttable}
\end{table*}

% \end{landscape}

\begin{landscape}
\begin{table}
\caption{RSG binary systems with their derived parameters. A full version of this table is available as supplementary material (online).}
\label{tb:params_bs}
\begin{tabular}{lcccccccccccc}
\hline
ID & F172M$^a$ & err & J$^b$ & Ks$^c$ & M$_1^d$ & $T_{\rm eff, 1}^{e}$ & $\log L_1/L_\odot^f$ & M$_2$ & err & $\log T_{\rm eff, 2}/K$ & $\log L_2/L_\odot$ & $\Delta RV$\\
& \multicolumn{2}{c}{ABmag} & mag & mag & M$_\odot$ & K & & \multicolumn{2}{c}{M$_\odot$} & & & \kms\\
\hline
SkKM  26 & 19.89 & 0.19 & 11.47 & 10.56 & 9.2 & 4290 & 4.26 & 4.1 & 0.6 & 4.25 & 2.59 & -- \\
PMMR  44 & 14.95 & 0.04 & 10.38 & 9.43 & 13.2 & 4220 & 4.71 & 13.5 & 2.2 & 4.43 & 4.617 & -- \\
LHA 115-S  14 & 18.09 & 0.13 & 10.44 & 9.44 & 13.1 & 4160 & 4.71 & 6.6 & 0.9 & 4.37 & 3.318 & 19.09 \\
SSTISAGEMA J005320.43-714448.8 & 19.74 & 0.2 & 12.31 & 11.4 & 7.3 & 4300 & 3.92 & 4.2 & 0.3 & 4.24 & 2.654 & -- \\
SkKM  87 & 17.35 & 0.11 & 10.62 & 9.66 & 12.1 & 4210 & 4.62 & 8.0 & 0.9 & 4.39 & 3.628 & -- \\
SSTISAGEMA J004145.95-733502.2 & 19.01 & 0.14 & 12.65 & 11.87 & 6.5 & 4500 & 3.73 & 4.9 & 0.4 & 4.26 & 2.937 & -- \\
SSTISAGEMA J004253.54-734028.8 & 20.26 & 0.27 & 12.57 & 11.67 & 6.9 & 4310 & 3.81 & 3.7 & 0.4 & 4.21 & 2.458 & -- \\
SSTISAGEMA J004445.26-732942.6 & 19.11 & 0.16 & 12.08 & 11.18 & 7.8 & 4310 & 4.01 & 4.9 & 0.6 & 4.27 & 2.896 & -- \\
LIN  82 & 17.04 & 0.06 & 11.69 & 10.69 & 8.9 & 4150 & 4.2 & 7.8 & 2.2 & 4.37 & 3.74 & 16.03 \\
SSTISAGEMA J004722.28-731613.0 & 18.79 & 0.14 & 11.25 & 10.32 & 9.8 & 4270 & 4.35 & 5.4 & 2.4 & 4.31 & 3.024 & -- \\
SSTISAGEMA J004415.16-731203.7 & 20.13 & 0.29 & 12.93 & 12.17 & 6.1 & 4520 & 3.62 & 3.7 & 0.6 & 4.21 & 2.511 & -- \\
BBB SMC 194 & 19.54 & 0.22 & 10.33 & 9.26 & 14.1 & 4040 & 4.78 & 4.6 & 2.4 & 4.28 & 2.724 & -- \\
SSTISAGEMA J004901.87-725405.0 & 19.19 & 0.17 & 11.95 & 11.13 & 7.9 & 4430 & 4.03 & 4.8 & 1.0 & 4.27 & 2.864 & -- \\
SSTISAGEMA J004640.16-730010.2 & 19.85 & 0.23 & 12.3 & 11.42 & 7.3 & 4350 & 3.91 & 4.1 & 0.6 & 4.23 & 2.613 & -- \\
2MASS J00451259-7311130 & 20.25 & 0.29 & 12.62 & 11.62 & 6.9 & 4150 & 3.83 & 3.7 & 1.2 & 4.21 & 2.461 & -- \\
SV* HV 11262 & 18.47 & 0.12 & 9.48 & 8.34 & 19.5 & 3930 & 5.15 & 6.1 & 3.9 & 4.36 & 3.166 & -- \\
SkKM  33 & 19.05 & 0.16 & 12.18 & 11.47 & 7.2 & 4600 & 3.9 & 4.9 & 1.3 & 4.27 & 2.92 & -- \\
SkKM  69 & 19.68 & 0.21 & 10.94 & 10.0 & 10.9 & 4250 & 4.48 & 4.4 & 0.5 & 4.27 & 2.671 & -- \\
SSTISAGEMA J005131.43-724621.0 & 18.67 & 0.13 & 11.43 & 10.5 & 9.3 & 4260 & 4.28 & 5.5 & 1.1 & 4.31 & 3.072 & -- \\
Dachs SMC 1-13 & 14.71 & 0.02 & 9.69 & 8.75 & 16.8 & 4250 & 4.98 & 15.4 & 2.9 & 4.45 & 4.73 & -- \\
SSTISAGEMA J004934.50-725252.9 & 18.61 & 0.2 & 11.9 & 11.1 & 8.0 & 4470 & 4.04 & 5.5 & 1.0 & 4.3 & 3.097 & -- \\
SkKM  84 & 19.86 & 0.24 & 10.49 & 9.46 & 13.0 & 4110 & 4.7 & 4.2 & 0.7 & 4.26 & 2.598 & -- \\
SkKM  62 & 19.04 & 0.27 & 10.94 & 10.05 & 10.7 & 4320 & 4.46 & 5.1 & 0.7 & 4.3 & 2.923 & -- \\
SSTISAGEMA J005357.99-721646.7 & 19.51 & 0.21 & 12.61 & 11.78 & 6.7 & 4410 & 3.77 & 4.4 & 0.5 & 4.24 & 2.745 & -- \\
SkKM  98 & 18.01 & 0.13 & 10.5 & 9.49 & 12.9 & 4140 & 4.68 & 6.8 & 1.3 & 4.37 & 3.353 & -- \\
PMMR  61 & 15.66 & 0.04 & 10.5 & 9.48 & 12.9 & 4130 & 4.69 & 11.6 & 2.2 & 4.43 & 4.332 & 19.56\\
SSTISAGEMA J005145.35-723114.8 & 16.39 & 0.04 & 12.67 & 11.83 & 6.6 & 4410 & 3.75 & 8.6 & 1.5 & 4.36 & 3.998 & -- \\
SSTISAGEMA J005207.63-723825.4 & 18.35 & 0.11 & 12.6 & 11.71 & 6.8 & 4330 & 3.8 & 5.6 & 0.9 & 4.29 & 3.198 & -- \\
PMMR  60 & 15.95 & 0.06 & 10.66 & 9.82 & 11.5 & 4410 & 4.55 & 10.5 & 2.2 & 4.42 & 4.203 & -- \\
SSTISAGEMA J005337.77-722519.7 & 18.68 & 0.14 & 12.19 & 11.37 & 7.4 & 4430 & 3.93 & 5.4 & 0.8 & 4.29 & 3.068 & -- \\
PMMR  57 & 19.32 & 0.2 & 11.1 & 10.22 & 10.1 & 4340 & 4.39 & 4.7 & 1.1 & 4.28 & 2.81 & -- \\
SkKM 145 & 17.07 & 0.06 & 11.56 & 10.67 & 8.9 & 4330 & 4.21 & 7.8 & 1.1 & 4.37 & 3.728 & -- \\
SSTISAGEMA J005516.33-720917.5 & 17.04 & 0.08 & 12.8 & 11.96 & 6.4 & 4390 & 3.7 & 7.2 & 0.9 & 4.33 & 3.727 & -- \\
SkKM 148 & 17.45 & 0.09 & 11.28 & 10.34 & 9.8 & 4240 & 4.35 & 7.4 & 0.8 & 4.36 & 3.576 & -- \\
SSTISAGEMA J005405.94-720849.2 & 19.39 & 0.18 & 11.63 & 10.66 & 8.9 & 4210 & 4.22 & 4.6 & 0.6 & 4.27 & 2.784 & -- \\
SSTISAGEMA J005446.17-721320.5 & 18.18 & 0.1 & 11.79 & 10.9 & 8.4 & 4330 & 4.12 & 6.1 & 0.6 & 4.32 & 3.268 & -- \\
SSTISAGEMA J005603.50-720157.2 & 19.59 & 0.2 & 12.35 & 11.46 & 7.2 & 4320 & 3.9 & 4.3 & 0.5 & 4.24 & 2.711 & -- \\
SkKM 112 & 19.03 & 0.21 & 10.56 & 9.61 & 12.4 & 4230 & 4.64 & 5.2 & 0.6 & 4.31 & 2.927 & -- \\
SSTISAGEMA J005630.86-714621.2 & 16.52 & 0.07 & 12.11 & 11.25 & 7.6 & 4370 & 3.98 & 8.3 & 0.9 & 4.36 & 3.949 & -- \\
SSTISAGEMA J005917.09-714836.9 & 18.91 & 0.13 & 12.07 & 11.19 & 7.8 & 4340 & 4.01 & 5.1 & 0.6 & 4.28 & 2.976 & -- \\
PMMR  70 & 19.37 & 0.18 & 9.32 & 8.36 & 19.3 & 4210 & 5.14 & 4.8 & 0.7 & 4.3 & 2.791 & -- \\
SSTISAGEMA J005526.02-724144.5 & 17.78 & 0.1 & 12.62 & 11.84 & 6.6 & 4500 & 3.74 & 6.1 & 0.9 & 4.31 & 3.429 & -- \\
SkKM 108 & 15.8 & 0.03 & 10.96 & 10.14 & 10.4 & 4440 & 4.43 & 10.3 & 1.6 & 4.41 & 4.26 & -- \\
SkKM 150 & 18.3 & 0.1 & 10.83 & 9.8 & 11.6 & 4110 & 4.56 & 6.2 & 1.3 & 4.35 & 3.226 & -- \\
\hline
\end{tabular}
\end{table}
\end{landscape}

\begin{landscape}
\begin{table}
\contcaption{RSG binary systems with their derived parameters.}
\label{tb:params_cont}
\begin{tabular}{lccccccccccccc}
\hline
ID & F172M$^a$ & err & J$^b$ & Ks$^c$ & M$_1^d$ & $T_{\rm eff, 1}^{e}$ & $\log L_1/L_\odot^f$ & M$_2$ & err & $\log T_{\rm eff, 2}/K$ & $\log L_2/L_\odot$ & $\Delta RV$\\
& \multicolumn{2}{c}{ABmag} & mag & mag & M$_\odot$ & K & & \multicolumn{2}{c}{M$_\odot$} & & & \kms\\
\hline
M2002 SMC  26342 & 17.42 & 0.07 & 11.8 & 10.83 & 8.5 & 4200 & 4.15 & 7.2 & 1.0 & 4.35 & 3.581 & -- \\
SSTISAGEMA J005733.89-720722.3 & 18.46 & 0.12 & 12.76 & 11.97 & 6.4 & 4480 & 3.69 & 5.4 & 0.6 & 4.28 & 3.156 & -- \\
PMMR  94 & 18.73 & 0.13 & 10.4 & 9.38 & 13.4 & 4130 & 4.73 & 5.6 & 0.7 & 4.33 & 3.051 & -- \\
SSTISAGEMA J010103.32-721838.7 & 19.25 & 0.18 & 12.52 & 11.75 & 6.7 & 4520 & 3.78 & 4.6 & 0.6 & 4.25 & 2.844 & -- \\
LHA 115-S  33 & 16.98 & 0.08 & 10.24 & 9.26 & 14.1 & 4180 & 4.78 & 8.9 & 0.8 & 4.42 & 3.795 & -- \\
LHA 115-S  30 & 17.08 & 0.06 & 8.97 & 7.98 & 22.3 & 4170 & 5.29 & 9.2 & 1.5 & 4.44 & 3.776 & 18.09\\
PMMR 118 & 16.92 & 0.05 & 10.44 & 9.41 & 13.3 & 4100 & 4.72 & 9.0 & 0.9 & 4.41 & 3.814 & 23.23\\
SkKM 202 & 16.79 & 0.05 & 10.89 & 9.93 & 11.1 & 4220 & 4.51 & 8.9 & 0.8 & 4.4 & 3.858 & -- \\
SSTISAGEMA J010108.96-721544.5 & 19.57 & 0.21 & 11.63 & 10.75 & 8.7 & 4340 & 4.18 & 4.4 & 0.7 & 4.26 & 2.716 & -- \\
PMMR 122 & 18.63 & 0.23 & 10.24 & 9.26 & 14.1 & 4180 & 4.78 & 5.8 & 0.7 & 4.34 & 3.096 & 14.79\\
Flo 288 & 19.64 & 0.28 & 10.93 & 9.95 & 11.0 & 4190 & 4.5 & 4.4 & 1.1 & 4.27 & 2.686 & -- \\
ISO-MCMS J005539.0-730850 & 18.76 & 0.24 & 11.18 & 10.2 & 10.2 & 4180 & 4.4 & 5.5 & 1.2 & 4.31 & 3.036 & -- \\
PMMR  74 & 20.08 & 0.26 & 9.74 & 8.88 & 16.2 & 4370 & 4.93 & 4.0 & 0.7 & 4.25 & 2.514 & -- \\
PMMR 124 & 14.52 & 0.02 & 10.38 & 9.38 & 13.4 & 4150 & 4.73 & 15.6 & 2.0 & 4.43 & 4.788 & -- \\
SkKM 172 & 19.76 & 0.27 & 10.53 & 9.51 & 12.8 & 4120 & 4.68 & 4.3 & 0.9 & 4.27 & 2.636 & -- \\
SSTISAGEMA J005911.99-723435.9 & 18.05 & 0.11 & 12.05 & 11.16 & 7.8 & 4320 & 4.02 & 6.2 & 0.9 & 4.32 & 3.32 & -- \\
PMMR 103 & 17.16 & 0.07 & 10.52 & 9.55 & 12.6 & 4190 & 4.66 & 8.4 & 0.9 & 4.4 & 3.71 & -- \\
SkKM 199 & 19.95 & 0.28 & 11.1 & 10.21 & 10.2 & 4320 & 4.4 & 4.0 & 0.6 & 4.25 & 2.567 & 16.17\\
SkKM 238 & 16.04 & 0.04 & 10.74 & 9.81 & 11.5 & 4260 & 4.56 & 10.4 & 0.8 & 4.42 & 4.167 & -- \\
M2002 SMC  54134 & 15.54 & 0.03 & 10.85 & 9.87 & 11.3 & 4190 & 4.54 & 11.2 & 1.5 & 4.42 & 4.371 & -- \\
SkKM 239 & 18.14 & 0.1 & 10.26 & 9.24 & 14.1 & 4130 & 4.78 & 6.6 & 1.1 & 4.37 & 3.301 & -- \\
Dachs SMC 2-10 & 16.56 & 0.07 & 10.24 & 9.24 & 14.1 & 4160 & 4.78 & 10.0 & 1.2 & 4.43 & 3.97 & -- \\
LHA 115-S  37 & 19.44 & 0.18 & 9.85 & 8.86 & 16.3 & 4170 & 4.94 & 4.7 & 0.5 & 4.29 & 2.764 & 21.99\\
PMMR 169 & 18.32 & 0.13 & 10.99 & 10.09 & 10.6 & 4320 & 4.44 & 6.1 & 0.7 & 4.34 & 3.217 & -- \\
SkKM 232 & 15.99 & 0.04 & 10.64 & 9.7 & 12.0 & 4250 & 4.6 & 10.6 & 1.2 & 4.42 & 4.19 & -- \\
SkKM 256 & 19.48 & 0.26 & 9.95 & 8.88 & 16.2 & 4040 & 4.93 & 4.6 & 0.8 & 4.29 & 2.748 & 12.05\\
SSTISAGEMA J010305.17-720917.1 & 19.92 & 0.28 & 12.08 & 11.27 & 7.6 & 4460 & 3.97 & 4.0 & 0.6 & 4.23 & 2.583 & -- \\
BBB SMC 376 & 19.92 & 0.25 & 10.65 & 9.77 & 11.7 & 4340 & 4.57 & 4.1 & 0.5 & 4.26 & 2.577 & -- \\
PMMR 149 & 16.17 & 0.04 & 10.57 & 9.62 & 12.3 & 4220 & 4.64 & 10.4 & 1.0 & 4.42 & 4.119 & -- \\
RMC  29 & 15.44 & 0.08 & 10.11 & 9.19 & 14.4 & 4280 & 4.81 & 12.7 & 1.8 & 4.44 & 4.424 & -- \\
Dachs SMC 2-27 & 15.98 & 0.05 & 10.58 & 9.61 & 12.4 & 4210 & 4.64 & 10.8 & 1.1 & 4.42 & 4.196 & -- \\
BBB SMC 307 & 17.84 & 0.09 & 9.81 & 8.81 & 16.5 & 4140 & 4.96 & 7.3 & 1.0 & 4.39 & 3.433 & -- \\
SSTISAGEMA J010339.88-723906.0 & 17.44 & 0.07 & 11.35 & 10.38 & 9.7 & 4210 & 4.33 & 7.4 & 0.8 & 4.36 & 3.581 & -- \\
BBB SMC 348 & 15.57 & 0.05 & 10.62 & 9.66 & 12.2 & 4210 & 4.62 & 11.4 & 1.1 & 4.42 & 4.364 & 15.67\\
LHA 115-S  50 & 17.81 & 0.12 & 10.87 & 9.92 & 11.1 & 4230 & 4.51 & 7.0 & 0.8 & 4.36 & 3.43 & -- \\
BBB SMC 231 & 18.15 & 0.14 & 9.8 & 8.79 & 16.6 & 4140 & 4.96 & 6.6 & 1.0 & 4.38 & 3.299 & -- \\
PMMR 178 & 18.86 & 0.21 & 10.14 & 9.19 & 14.4 & 4230 & 4.8 & 5.4 & 0.6 & 4.33 & 3.0 & -- \\
BBB SMC 115 & 17.76 & 0.12 & 11.0 & 10.13 & 10.4 & 4360 & 4.43 & 7.0 & 0.7 & 4.36 & 3.452 & -- \\
BBB SMC 229 & 19.13 & 0.23 & 11.86 & 10.95 & 8.3 & 4290 & 4.1 & 4.9 & 0.4 & 4.28 & 2.889 & -- \\
SSTISAGEMA J012150.98-733740.9 & 16.89 & 0.07 & 12.58 & 11.75 & 6.7 & 4420 & 3.78 & 7.5 & 0.8 & 4.34 & 3.789 & -- \\
SSTISAGEMA J012138.30-732253.2 & 18.52 & 0.14 & 12.82 & 12.09 & 6.2 & 4580 & 3.65 & 5.3 & 0.3 & 4.27 & 3.133 & -- \\
SSTISAGEMA J012651.76-732100.6 & 17.62 & 0.1 & 12.36 & 11.62 & 6.9 & 4560 & 3.83 & 6.4 & 0.5 & 4.32 & 3.493 & -- \\
Gaia EDR3 4689009311093195136 & 19.53 & 0.21 & 11.36 & 10.53 & 9.2 & 4420 & 4.27 & 4.5 & 0.5 & 4.27 & 2.731 & -- \\
Gaia EDR3 4688998732646783232 & 19.96 & 0.24 & 12.05 & 11.08 & 8.0 & 4200 & 4.05 & 4.0 & 0.5 & 4.23 & 2.566 & -- \\
\hline
\end{tabular}
\end{table}
\end{landscape}

\begin{landscape}
\begin{table}
\contcaption{RSG binary systems with their derived parameters.}
\label{tb:params_cont2}
\begin{tabular}{lcccccccccccccc}

\hline
\multicolumn{12}{l}{\footnotesize{$^a$ Photometry from Thilker et al. (in prep.)}}\\
\multicolumn{12}{l}{\footnotesize{$^b$ Photometry from The Two Micron All Sky Survey~\citep{2006AJ....131.1163S}. Uncertainties in range 0.020 to 0.030.}}\\
\multicolumn{12}{l}{\footnotesize{$^c$ Photometry from The Two Micron All Sky Survey~\citep{2006AJ....131.1163S}. Uncertainties in range 0.016 to 0.030.}}\\
\multicolumn{12}{l}{\footnotesize{$^d$ Typical uncertainties of $\pm 1.5$. Uncertainties listed in online version of table}}\\
\multicolumn{12}{l}{\footnotesize{$^e$ Uncertainties of $\pm 150$\,K.}}\\
\multicolumn{12}{l}{\footnotesize{$^f$ Typical uncertainties of $\pm 0.12$.}}\\
\end{tabular}
\end{table}
\end{landscape}

\section{SMC RSG effective temperatures}            \label{ap:Teff}

TDN18 highlighted a potential systematic offset between their results and those of~\citet{2015ApJ...806...21D} in the SMC.
This is expanded upon by \citet{2021MNRAS.505.4422G} who re-determined the effective temperatures for the sample of~\citep{2013ApJ...767....3D,2015ApJ...806...21D}, and found excellent agreement with the results of~\citet{2013ApJ...767....3D} and~\citet{2015ApJ...806...21D}.
\citet{2021MNRAS.505.4422G} found a systematic offset of $\sim$ 150\,K between their results and those of TDN18, with the TDN18 temperatures being the cooler.
\citet{2021MNRAS.505.4422G} explain these differences as the result of a combination of not accounting for non-LTE effects~\citep[see][]{2013ApJ...764..115B} and continuum placement issues in the wavelength range studied by TDN18.

\citet{2021ApJ...922..177M} determined effective temperatures of RSGs in the SMC using de-reddened $J-K$ colours, assuming an $A_{\rm V}=0.75$\,mag and the reddening law of~\citet{1998ApJ...500..525S}.
\citet{2021ApJ...922..177M} calibrated the $J-K$ colours in the SMC using an extension to the MARCS one-dimensional atmospheric models, as described in~\citet{Levesque06}.
These authors noted that the calibration of $J-K$ colours to the MARCS atmospheric models to determine RSG effective temperatures requires an additional offset of 200\,K to better match the~\citet{Levesque05, Levesque06} RSG temperature scales, which is determined using spectral fitting to the broad TiO bands at optical wavelengths.
This offset is in the sense that the $J-K$ effective temperatures should be shifted to cooler temperatures~\citep{2021ApJ...922..177M}.

To better understand the differences between these alternative methods to determine RSG effective temperatures we directly compare the relationships used by~\citet{2021MNRAS.502.4890D} and~\citet{2021ApJ...922..177M}.
We find very good agreement between the two relations with an average offset of 55\,K with a standard deviation of 15\,K over a range of 3500 to 4500\,K.

We choose to implement neither the +150\,K offset suggested by~\citet{2021MNRAS.505.4422G} nor the $-$200\,K suggested by~\citet{2021ApJ...922..177M}.

%%%%%%%%%%%%%%%%%%%%%%%%%%%%%%%%%%%%%%%%%%%%%%%%%%

% Don't change these lines
\bsp	% typesetting comment
\label{lastpage}
\end{document}